\documentclass[prl, twocolumn]{revtex4}
\usepackage{graphicx}
\usepackage{amssymb}

\begin{document}

\title{Dynamical Casimir effect in a Josephson metamaterial}

\author{Pasi L\"ahteenm\"aki$^1$, G. S. Paraoanu$^1$, Juha Hassel$^2$, and Pertti J. Hakonen$^1$}

\affiliation{$^1$Low Temperature Laboratory, Aalto University, PO Box 15100, FI-00076 AALTO,
Finland}

\affiliation{$^2$VTT Technical Research Centre of Finland, PO BOX 1000, FI-02044 VTT, Finland}

\maketitle
\small
\textbf{
Vacuum modes confined into an electromagnetic cavity give rise to an attractive interaction between the 
opposite walls \cite{casimir}. When the distance between the walls is changed non-adiabatically, virtual 
vacuum modes are turned into real particles, i.e. photons are generated out of the vacuum 
\cite{moore,fullingdavies}. These effects are known as the static and dynamical Casimir effect, 
respectively. Here we demonstrate the dynamical Casimir effect using a Josephson metamaterial embedded in 
a microwave cavity at 5.4 GHz. We achieve the non-adiabatic change in the effective length of the cavity 
by flux-modulation of the SQUID-based metamaterial, which results in a few percent variation in the 
velocity of light. We show that energy-correlated  photons are 
generated from the ground state of the cavity and that their power spectra display a bimodal frequency 
distribution. These results are in excellent agreement with theoretical predictions, all the way to the 
regime where classical parametric effects cannot be of consequence.
}

A fundamental theoretical result of modern quantum field theory is that
the quantum vacuum is unstable under certain external perturbations that
produce otherwise no consequences in a classical treatment \cite{first}.
As a
result of this instability, virtual fluctuations populating the quantum vacuum
are converted into real particles by the energy provided by the
perturbation.
For example, the application of an intense electric fields extracts
electron-positron pairs from vacuum (Schwinger effect), the bending of
space-time at in the intense gravitational field at event horizons
results in black hole evaporation (Hawking radiation), the acceleration
of an observer in the Minkowski vacuum results in the detection of
particles (Unruh effect), and sudden changes in the boundary conditions
of electromagnetic field modes or in the equations of motion creates photons (dynamical Casimir
effect) \cite{review}.
None of these dynamical effects have been yet experimentally verified, although preliminary evidence for the analog of
Hawking radiation and dynamical Casimir effect have very recently been reported \cite{hawking,delsing}.

In general, the dynamical Casimir effect (DCE) is the phenomenon of particle creation from vacuum due to modulation of the background in which a quantum field propagates \cite{yablo}. There are several proposals how to observe these phenomena using single vibrating mirrors or vibrating cavities \cite{lambrecht,dodonovPRA,jiPRA,schutzholdPRA}.
In addition, several alternative proposals for experiments that could lead to a measurable
flux of Casimir photons exist, for example sonoluminescence \cite{schwinger}, changing the length of three-dimensional cavity resonators by using a
laser-modulated reflectivity of semiconducting layers \cite{semi1,schutzholdPRL,semi2}, surface acoustic waves
\cite{saw}, and superconducting qubits \cite{ciuti}. The experimental verification of the dynamical Casimir
effect in mechanically modulated systems has so far been hampered by the fast speeds required for the change in order to obtain a measurable
amount of energy \cite{review}. A moving mirror creates photons out of vacuum fluctuations in pairs whose
frequencies sum up to the pump frequency of the mirror. Typically, the photon production is non-negligible
only when the mirror velocity approaches the speed of light. For example, a mirror on a 1 GHz
nanomechanical oscillator moving at an amplitude of 1 nm would create only about $10^{-9}$ Casimir
photons/s. Recently, it was suggested that these speeds can be obtained in a circuit QED (quantum electrodynamics)
setup in which the boundary condition is realized by a SQUID (superconducting quantum interference device)
that terminates a superconducting waveguide \cite{ChalmersPRL,ChalmersPRA,photongener}.

\begin{figure}[h]
\begin{center}
\includegraphics[width=8cm]{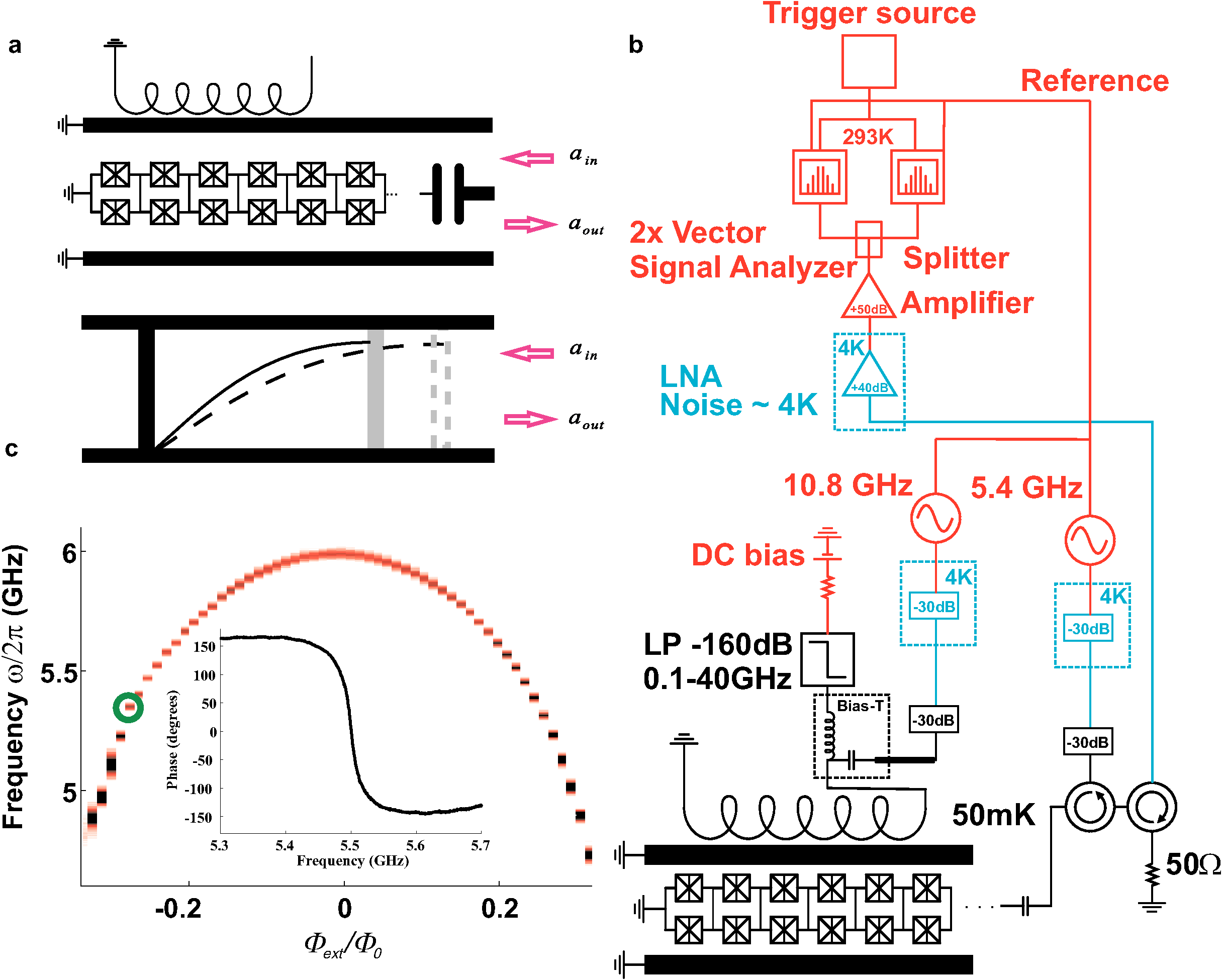}
\end{center}
\caption{(a) Equivalent electrical and mechanical circuits: the modulation of the Josephson inductance in
the metamaterial by magnetic flux $\Phi$ varies $\lambda/d$, the wave length with respect to the cavity length,
which is equivalent to modulating the effective length $d$ of the cavity by
mechanical means. The coupling capacitor is equivalent to a semitransparent mirror.
(b) Schematics of the measurement setup.
The metamaterial sample is a 7-mm-long coplanar waveguide with 250 embedded SQUIDs, each junction
having a critical current of $\sim 10$ $\mu$A. Modulation of $\Phi$ through the SQUIDs is applied using a lithographically fabricated spiral coil underneath the metamaterial. c) Resonant frequency $\omega_{\rm res}/2\pi$ vs.
reduced magnetic flux $\Phi_{\rm ext}/\Phi_0$ without the pump signal; the DC operating point for DCE experiments is denoted by a green circle. The inset displays the measured phase of the
scattering parameter $S_{11}$ while sweeping $\Phi$. The steepness of the variation in the phase $\arg(S_{11})$
governs the effective "movement of the mirrors".
}
\label{schematic}
\end{figure}

In our work, we use a metamaterial consisting
of an array of 250 SQUIDs, embedded into a 7 mm long superconducting coplanar waveguide.
In order to enhance the photon production rate even further, we
operate the SQUID array as a low $Q_{\rm res}$ resonator ($Q_{\rm res} \sim 100$) by coupling the metamaterial through a capacitor
to the external transmission line (see Figs. \ref{schematic}a and \ref{schematic}b). The flux-tunable SQUIDs allow us to
modulate the speed of light ($v\approx0.5 c_0$, with $c_0$ the speed in vacuum) and thus change the effective electrical length of the cavity at drive frequencies $\omega_d$ on the order of 10 GHz over a relatively large range, up to 10 mm.
Therefore, the field inside the cavity experiences an effective modulation of boundary conditions (see Fig. 1a).
 Moreover, the input-output theory \cite{walls, RMP} predicts that the field reflected by the cavity acquires a phase that depends on the detuning with respect to the resonant frequency of the cavity. From the point of view of an observer detecting the output field, this change of phase could have come as well from an effective mirror reflecting the input field or from an other appropriately chosen boundary condition. The modulation of the resonance frequency is thus analogous to a modulation of such a boundary condition.
The detailed arguments of this equivalence are outlined in the Supplementary material.
The resonance conditions facilitate the
creation of a large amount of radiation even at temperatures where the field modes are essentially
unoccupied, and we detect a clear correlation at two frequencies symmetric around half the drive frequency $\omega_{d}/2$.
We observe two-mode squeezed (TMS) photons and a characteristic double peak
\cite{ChalmersPRA} in the squeezed output noise spectrum when the resonator is detuned from $\omega_{d}/2$.

The samples were fabricated using a NbAlO$_x$Nb trilayer process. The metamaterial cavity was cooled down to
$T=50$ mK in a pulse-tube-based dilution refrigerator; $T$ corresponds to a thermal occupation number of $\bar{n} = 0.0056$ at 5.4
GHz. All the lines used to flux bias the sample and to feed the microwave signals are strongly attenuated:
the microwave attenuation $\gamma > 160$ dB in the DC-feed and, for the microwave cables, $\gamma \sim 70$
dB to guarantee proper thermalization of the background radiation. The connection to the low noise
amplifier (LNA) is isolated by two circulators in combination with high- and low-pass filters at the base
temperature, providing isolation on the order of 40 dB from the 4 K stage.

We first established that the device worked indeed as a cavity whose
effective length is controlled by an external magnetic field. Fig. \ref{schematic}c shows the variation of
$\omega_{\rm res}/2\pi$ with magnetic flux bias $\Phi$.
The inset displays the measured phase of the
reflection coefficient $S_{11}$ which changes steeply while sweeping $\Phi$. The variation of $\arg(S_{11})$
is a direct measure of how much the wavelength $\lambda \simeq 4d$ changes with the  external flux; here $d$ denotes the physical length of the resonator.
The change in $\lambda/d$ due to modulation is the equivalent to the "movement of the mirror". As shown in the Supplementary material, the resonance quality factor enhances the length modulation which becomes $
\left. \delta L^{\rm (eff)}_{\omega} \right|_{\omega = \omega_{\rm res}}= Q_{\rm res}\frac{c}{\omega_{\rm res}}\frac{\delta l}{l}$,
where $l$ denotes the inductance per unit length and $\delta l$ its change due to the modulation.

We also checked that our metamaterial works as a phase-sensitive parametric amplifier \cite{yurke,lehnert,eichler} in the three wave mode
when pumped at $\omega_d=2\omega_{\rm res}$. At a pump power $P_d = -75$ dBm (after the low-T attenuators), the
gain in the degenerate mode reached 23 dB and noise temperature at 5.5 GHz became $T_n = 0.4 \pm 0.2$ K as
deduced from the increase in the signal/noise ratio. Taking into account the losses in the circuitry of the
S/N ratio measurement, this result indicates that the metamaterial behaves as designed.

\begin{figure}[h]
    \begin{center}
    \includegraphics[width=8cm]{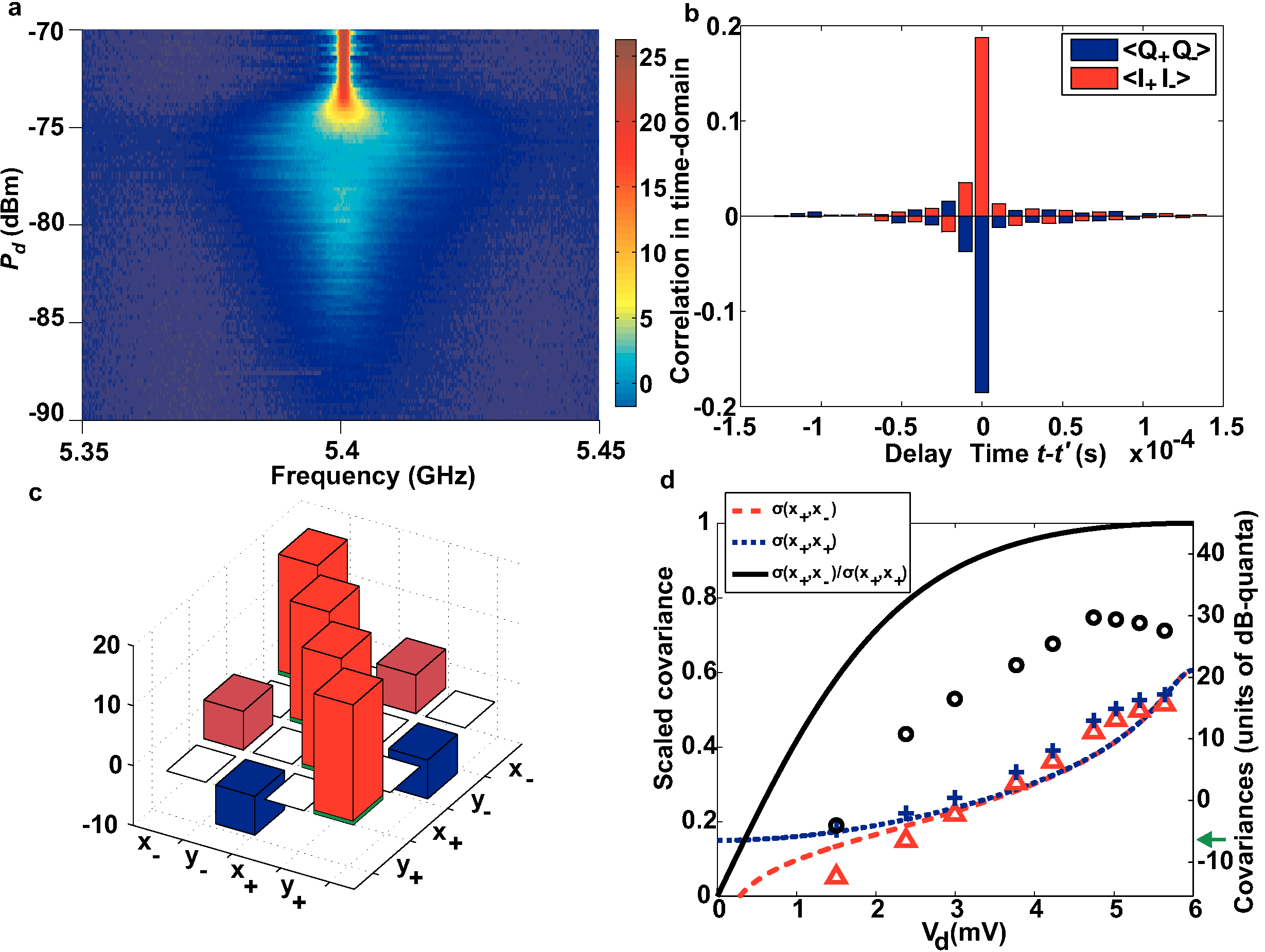}
	\caption{
(a) Color map of the output noise power (color scale in dB units with $ k_B \times 5\textrm{ K} $ as the reference level) measured across the resonant
frequency $\omega_{\rm res}/2\pi=5.4$ GHz while increasing the pump
power $P_d$ at $\omega_{d}=10.8$ GHz.  At $P_d >-75$ dBm, the device becomes unstable against self-sustained oscillations.
(b) Cross-correlations for in-phase ($I$) and quadrature ($Q$) amplitudes $\langle I_+(t) I_-(t') \rangle$ and $\langle Q_+(t) Q_-(t') \rangle$
between sidebands at $\omega_{\pm}=\omega_{d}/2 \pm \nu $ vs. time delay $t-t'$ with the pump
on. The pump is at 10.8 GHz while the IQ-demodulators are at 5.398 GHz and 5.402 GHz. The used bandwidth is
200 kHz and $10^6$ samples were collected.
(c) Covariance matrix $\sigma(A,B)$, measured at $P_d=-74$ dBm. The diagonal elements amount to $0.5 \cdot {\rm DCE} + \frac{1}{4}$ quanta (marked in red and green, respectively),
while the non-zero off-diagonal elements describe the squeezing correlations of the system.
(d) Covariances  $\sigma(x_+,x_-)$ ($\Delta$), $ \sigma(x_+,x_+)$ (+), and their ratio $\sigma(x_+,x_-)/\sigma(x_+,x_+)$ (o)
vs. modulation power $P_d$ in the flux coil; the ratio and its fit are on linear scale (left), whereas the rest is on logarithmic scale (dB, right).
The dashed and dotted curves represent theory predictions, obtained using Eqs. (S50-S53), (S56-S58), respectively, and the solid curve is their ratio. On the right scale, $\frac{1}{4}$ of a quantum corresponds to -6 dB (denoted by arrow).
The parameter values used in the evaluation are listed in the caption of Fig. 3.
} \label{corr}
    \end{center}
\end{figure}

The existence of the Casimir photon generation was investigated at temperatures $T=50 - 230$ mK
by measuring the noise power spectra (see Fig. 2a) when the flux was modulated around a fixed DC bias point
at twice the resonant frequency $\omega_{\rm res}/2\pi$.
To understand these results, we present first a simplified theoretical
description of the Casimir effect in our system.

One can intuitively picture the dynamical Casimir effect as follows: mathematically, the metamaterial is
equivalent to a gravitational pendulum with length proportional to the total Josephson energy $E_{\rm J}$.
The mass of the pendulum is the analog of the electrical capacitance of the junction and the angle with
respect to the vertical is the analog of the superconducting phase difference. By modulating the effective
length at twice the natural frequency of the pendulum, any initial oscillation or fluctuations of the
pendulum become parametrically amplified. However, in the absence of any initial oscillation, changes in
the length of the pendulum do not produce any oscillation, according to classical physics. Yet the
quantum-mechanical prediction is different: due to the uncertainty principle, the pendulum is never at
rest, and the small quantum fluctuations of the ground state become parametrically
transformed into real, measurable oscillations of the superconducting phase. According to the Josephson relations, these
oscillations of the phase produce an oscillatory voltage between the signal line and the ground of our
transmission line, which propagates as real photons in the transmission line. Hence, the Josephson
metamaterial acts as an antenna broadcasting information about the local vacuum. The mathematical analysis
of the DCE in a Josephson metamaterial is presented in the Supplementary material. The flux modulation
leads to coupling of modes $a (\frac{\omega_d}{2} +\nu )$ and $\left[ a (\frac{\omega_d}{2} -\nu )\right]^{\dagger}$
(see Eqs. (S26) and (S49)).

Our experimental results on the dynamic Casimir photon generation in Fig. \ref{corr}a display a
strong increase in noise power around $\omega_{\rm res}
$ when $P_d$ at $\omega_d/2\pi$ is enhanced.
The increase in the noise level is abruptly saturated when a parametric
oscillation sets in at $P_d=-75$ dBm. Below the threshold, the emitted power spectrum follows closely the predicted form for a lossless cavity
\begin{equation}
{\rm DCE}(\nu )= \frac{\kappa^{2}|\alpha|^{2}}{|{\cal N}(\nu)|^{2}}
\left|\chi \left(\frac{\omega_{d}}{2}+\nu \right)\right|^{2}
\left|\chi \left(\frac{\omega_{d}}{2}-\nu \right)\right|^{2}, \label{DCEideal}
\end{equation}
where ${1/|\cal N}(\nu)|^{2}$ is a factor producing parametric gain (in S26), $\chi \left(\omega \right)$ is the electrical susceptibility of the resonator with $\kappa$ as the cavity decay rate (S37), and $\alpha$ denotes the magnitude of the flux pumping (S14). The theoretical model can be
extended/generalized to account for the losses in the cavity
(see Eqs. (S50-S53)).

The Casimir photons are generated as correlated pairs with frequencies $\omega_{\pm} = \omega_d/2 \pm \nu$, with energy conservation satisfied $\omega_{+}+\omega_{-} = \omega_d$. To study experimentally the correlations between the sidebands $\omega_{\pm}$ , we use two vector signal analyzers triggered simultaneously to collect in-phase and out-of-phase time-domain IQ-waveforms at $\omega_{\pm}$ and with homodyning phases $\theta_{\pm}$. We calculate the correlations between the IQ data and eliminate the amplifier noise by a procedure described in detail in the Supplementary material. The resulting correlations, which we denote by $\langle I_{+} I_{-} \rangle$, $\langle I_{+} Q_{-} \rangle$, {\it etc.}, turn out to be directly related to the correlations of the quadratures $x_{\pm} = (1/2)(a_{\pm} + a_{\pm}^{\dag})$ and $y_{\pm} = (1/2i)(a_{\pm} - a_{\pm}^{\dag})$ of the outgoing field $\{a_{\pm}, a_{\pm}^{\dag} \}$ in the two sidebands. For convenience, the homodyning phases $\theta_{+}$ and $\theta_{-}$ of the two analyzers and the phase of the pump field have been rotated so that $\langle x_{\pm}y_{\mp}\rangle = \langle y_{\pm}x_{\mp}\rangle = 0 $. We have verified that all the measured quadrature correlations behave according to the theory under this rotation. Since this procedure is performed for every point at which we measure correlations, there is no need write down the phases $\theta_{\pm}$ any more. To put the correlations in a compact form, we introduce the covariance matrix $\sigma (A,B) = \frac{1}{2} \langle \{A,B \}\rangle$, where $\{ ,\}$ denotes the anticommutator and  $A$ and $B$ are any of the operators $x_{\pm}$, $y_{\pm}$. The experimental results for correlations are displayed in Fig. \ref{corr}. The time dependence of $\langle I_+I_-\rangle$ and $\langle Q_+Q_-\rangle$ is displayed in Fig.
\ref{corr}b, and our results for the full correlation matrix are illustrated in Fig. \ref{corr}c.

In order to compare the theoretical predictions for Casimir photon generation with our results, Fig. 2d displays the pumping power dependence of the correlators
 $\sigma(x_+,x_-)$, $ \sigma(x_+,x_+)$, and their ratio $\sigma(x_+,x_-)/\sigma(x_+,x_+)$. When compared with theory, $\sigma(x_+,x_+)$ follows closely $0.5 \cdot {\rm DCE} + \frac{1}{4}$ (dotted curve) and the covariance $\sigma(x_+,x_-)$ agrees with the calculated two-mode correlations (dashed curve). The scaled squeezing correlator $\sigma(x_+,x_-)/\sigma(x_+,x_+)$ displays the correct linear dependence on the pumping drive amplitude $V_d$ but its magnitude falls a bit below the theoretical expectation.

\begin{figure}[h]
    \begin{center}
     \hspace{-0.3cm}
     \includegraphics[width=8cm]{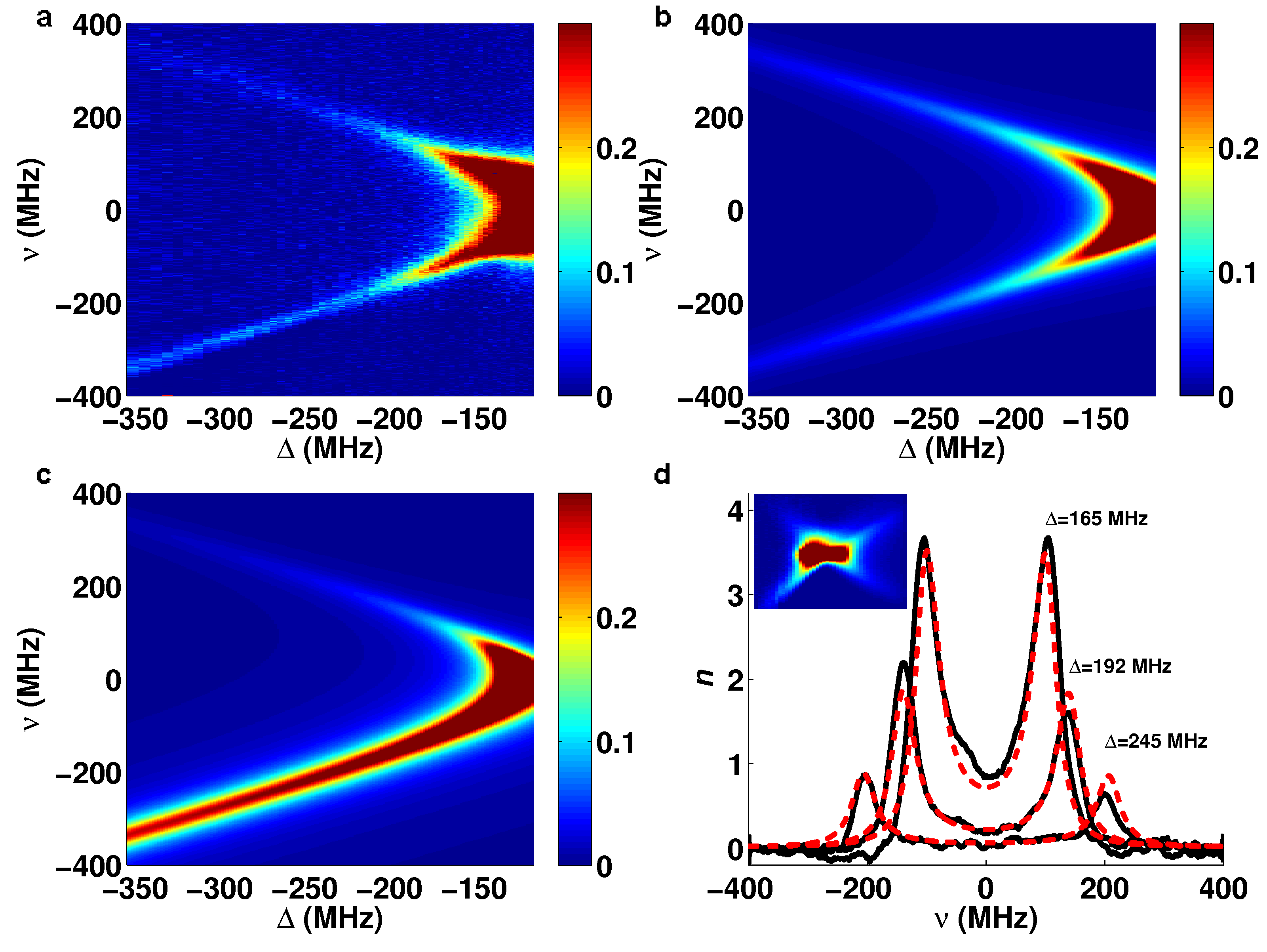}
     \caption{
a) Emitted noise as a function of $\nu = \omega - \omega_d/2$ while tuning
$\Delta = \omega - \omega_d/2$ measured around $\omega_d/2$=5.4 GHz at 50 mK.
b) Theoretical prediction form Eqs. (S50-S53)
evaluated using parameters $\alpha=2 \pi \times 65 \textrm{ MHz}, \kappa_{\rm E}=2 \pi \times 15 \textrm{ MHz} \textrm{ (at 50  mK)}, \kappa_{\rm I}=2 \pi \times 9.2 \textrm{ MHz} \textrm{ (at 60  mK)}$;
in both (a) and (b), the color scale for the noise power is cut at 0.3 dB in the vicinity of $\omega_d/2$.
c) Theoretical prediction of classical vacuum calculated at a single quantum of classical noise in the 
internal modes.
d) Slices with constant detuning $\Delta$ taken from (b) are compared with the
theoretical predictions from Eqs. (S50-S53); parameter values as above. The inset depicts the full, symmetric noise pattern measured at smaller pump
power over the range $-100 < \Delta <100$ MHz and  $-100 < \nu <100$ MHz.
\label{fig006}}
    \end{center}
\end{figure}


Finally, we have measured what happens when the resonator is detuned from $\omega_d/2$ while keeping
$P_d={\rm const}$. As seen in Fig. 3a, we observe a splitting of the noise spectrum into two peaks, which
are in good agreement with our theoretical results depicted in Fig. 3b at $T_{\rm E}=50$ mK and
$T_{\rm I}=60$ mK for the external ($\kappa_{\rm E}$) and internal ($\kappa_{\rm I}$) dissipation temperature, respectively.
A classical description of the measured spectra as resulting from uncontrollable additional sources of 
noise would utterly fail to explain the structure presented in Fig. 3a. As an example, in Fig. 3c we show 
the effect of a single quantum of classical noise in the internal modes: the spectrum becomes strongly 
asymmetric \cite{note}, due to the asymmetric dependence of the emitted cavity power on the internal 
temperature in the "lower arm" in Figs. 3a and 3b (see Eqs. S50-53). Thus, the symmetric appearance of the 
pattern in Fig. 3a can be regarded as an additional signature of the quantum nature of the photon 
generation in our system.

The experimental results at three different values of detuning are plotted and compared with the theoretical
predictions in Fig. 3d. Based on these fits, or more precisely from the symmetry of the lower
and upper peaks, we can determine that the metamaterial
temperature is not much different from the measured $T = 50$ mK. 
The observed bimodal structure of the spectrum with nearly equally strong peaks at $\nu \sim
\Delta$ can be considered as direct evidence of the existence of the dynamical Casimir effect.
Furthermore, these observations were made in a regime where the parametric
gain of the driven cavity is nearly unity, in clear distinction to regular parametric amplifiers and
optical parametric oscillators producing squeezed states. Our results pave the way for EPR-type of experiments with correlated microwave photon pairs and for quantum information processing with continuous variables \cite{ou,braunstein}.

%
%
%


\begin{thebibliography}{99}

\bibitem{casimir} Casimir, H. B. G.,  On the attraction between two perfectly conducting plates. \emph{Koninkl. Ned. Adak. Wetenschap. Proc.} \textbf{51}, 793 (1948).

\bibitem{moore} Moore, G. T., Quantum theory of the electromagnetic
field in a variable-length one-dimensional cavity.
{\it J. Math. Phys.} {\bf 11}, 2679--2691 (1970).

\bibitem{fullingdavies} Fulling, S. A., \& Davies, P. C. W., Radiation from a
Moving Mirror in Two Dimensional Space-Time: Conformal Anomaly.
{\it Proc. R. Soc. London}, Ser. {\bf A 348}, 393--414 (1976).

\bibitem{first} For a review, see Fulling, S. A., Aspects of quantum field theory in curved space-time
(Cambridge Univ. Press, Cambridge, 1989).

\bibitem{review} See,\emph{ e.g}., Nation, P.D., Johansson, J. R., Blencowe, M. P., \& Nori, F.,
Stimulating Uncertainty: Amplifying the Quantum Vacuum with Superconducting Circuits.
arXiv:1103.0835v2 (\emph{Rev. Mod. Phys.}, in press 2011) and references therein.

\bibitem{delsing} Wilson, C. M., Johansson, G., Pourkabirian, A., Simoen, M., Johansson, J. R., Duty, T., Nori, F. \& Delsing, P., Observation of the dynamical Casimir effect in a superconducting circuit. ArXiv:1105.4714.

\bibitem{hawking} Belgiorno, F., Cacciatori, S. L., Clerici. M., Gorini, V., Ortenzi, G., Rizzi, L.,
Rubino, E., Sala, V. G., \& Faccio, D.,
Hawking Radiation from Ultrashort Laser Pulse Filaments.
{\it Phys. Rev. Lett.} {\bf 105}, 203901-4 (2010).

\bibitem{yablo} Yablonovitch, E.,
Accelerating Reference Frame for Electromagnetic Waves in a Rapidly Growing Plasma:
Unruh-Davies-Fulling-DeWitt Radiation and the Nonadiabatic Casimir Elffect,
\emph{Phys. Rev. Lett.} \textbf{62}, 1742-1745 (1989).

\bibitem{lambrecht} Lambrecht, A., Jaekel, M.T., \& Reynaud, S.,
Motion Induced Radiation from a Vibrating Cavity.
{\it Phys. Rev. Lett.} {\bf 77}, 615-618 (1996).

\bibitem{dodonovPRA} Dodonov, V. V. \& Klimov, A. B.,
Generation and detection of photons in a cavity with a resonantly oscillating boundary.
{\it Phys. Rev. A} {\bf 53}, 2664-2682 (1996).

\bibitem{jiPRA} Ji, J.-Y., Hyun-Hee Jung, H.-H., Park, J.-W. \& Soh, K.-S.,
Production of photons by the parametric resonance in the dynamical casimir effect.
{\it Phys. Rev. A} {\bf 56}, 4440-4444 (1997).

\bibitem{schutzholdPRA} Sch\"utzhold, R., Plunien, G. \& Soff, G.,
Trembling cavities in the canonical approach,
{\it Phys. Rev. A} {\bf 57}, 2311-2318 (1998).


\bibitem{schwinger} Schwinger, J., Casimir energy for dielectrics, \emph{Proc. Nat. Acad. Sci.}\textbf{ 89}, 4091-4093 (1993).

\bibitem{schutzholdPRL} Uhlmann, M., Plunien, G., Sch\"utzhold, R., \& Soff, G., Resonant cavity photon
creation via the dynamical Casimir effect,
{\it Phys. Rev. Lett.} {\bf 93}, 193601-4 (2004).

\bibitem{semi1} Crocce, M., Dalvit, D.A.R., Lombardo, F. C. \& Mazitelli, F. D.,
Model for resonant photon creation in a cavity with time-dependent conductivity.
\emph{Phys. Rev. A} {\bf 70}, 033811-6 (2004).

\bibitem{semi2} Braggio, C., Bressi, G., Carugno, G., Del Noche, C., Galeazzi, G.,
Lombardi, A., Palmieri, A., Ruoso, G. \& Zanello, D.,
A novel experimental approach for the detection of the dynamical Casimir effect.
{\it Europhys. Lett.} {\bf 70}, 754-760 (2005).

\bibitem{saw} Kim, W.-J., Brownell, J. H. \& Onofrio, R.,
Detectability of dissipative motion in quantum vacuum via superradiance.
{\it Phys. Rev. Lett.} {\bf 96}, 200402-5 (2006).

\bibitem{ciuti} De Liberato, S., Gerace, D., Carusotto, I. \& Ciuti, C.,
Extracavity quantum vacuum radiation from a single qubit.
\emph{Phys. Rev. A }{\bf 80}, 053810-1/5 (2009).

\bibitem{dodonov} Dodonov, V. V.,
Current status of the Dynamical Casimir Effect.
{\it Physica Scripta} {\bf 82}, 038105-14 (2010).

\bibitem{ChalmersPRL} Johansson, J. R., Johansson, G., Wilson, C. M. \&  Nori, F.,
Dynamical Casimir Effect in a Superconducting Coplanar Waveguide.
{\it Phys. Rev. Lett.} {\bf 103}, 147003-6 (2009).

\bibitem{ChalmersPRA} Johansson, J. R., Johansson, G., Wilson, C. M. \&  Nori, F.,
Dynamical Casimir effect in superconducting microwave circuits.
{\it Phys. Rev. A} {\bf 82}, 052509-17 (2010).

\bibitem{photongener} Wilson, C. M., Duty, T., Sandberg, M., Persson, F., Shumeiko, V. \& Delsing, P.,
Photon Generation in an Electromagnetic Cavity with a Time-Dependent Boundary. \emph{Phys. Rev. Lett.} \textbf{105}, 233907-10 (2010).

\bibitem{walls} Walls D. F. \& Milburn G. J., "Quantum Optics", 2nd edition, Springer (2008).

\bibitem{RMP} Clerk A. A., Devoret M. H., Girvin S. M., Marquardt F. \& Schoelkopf R. J.,
Introduction to Quantum Noise, Measurement and Amplification.
{\it Rev. Mod. Phys.} {\bf 82}, 1155-1208 (2010).

\bibitem{yurke}
Yurke, B., Corruccini, L. R., Kaminsky, P. G., Rupp, L. W.,
 Smith, A. D., Silver, A. H., Simon, R. W. \& Whittaker, E. A.,
Observation of parametric amplification and deamplification.
in a Josephson parametric amplifier, \emph{Phys. Rev. A} \textbf{39}, 2519-2533 (1989).

\bibitem{lehnert}
Castellanos-Beltran, M. A., Irwin, K. D., Hilton, G. C., Vale, L. R. \& Lehnert, K. W.,
Amplification and squeezing of quantum noise with a tunable Josephson metamaterial.
\emph{Nat. Phys.} \textbf{4}, 928-931 (2008).

\bibitem{eichler}
Eichler, C., Bozyigit, D., Lang, C.,  Baur, M., Steffen, L., Fink, J. M., Filipp, S. \& Wallraff, A.,
Observation of Two-Mode Squeezing in the Microwave Frequency Domain.
\emph{Phys. Rev. Lett.} \textbf{107}, 113601-4 (2011).


\bibitem{caves85} Caves C. M. \&  Schumaker B. L.,
New formalism for two-photon quantum optics. I. Quadrature phases and squeezed states.
\emph{Phys. Rev. A} {\bf 31}, 3068-3092 (1985).

\bibitem{ou}Z. Y. Ou, S. F. Pereira, H. J. Kimble, and K. C. Peng,
Realization of the Einstein-Podolsky-Rosen paradox for continuous variables,
Phys. Rev. Lett. \textbf{68}, 3663–3666 (1992).

\bibitem{braunstein}
Braunstein, S. L. \& van Loock, P.,
Quantum information with continuous variables.
\emph{Rev. Mod. Phys.} \textbf{77}, 513-575 (2005).

\bibitem{note} A similar pattern is obtained classically by a
time-domain simulation of a parallel LC tank circuit with resistance,
when the inductance is pumped at $2\omega_d$ and the resistor acts as the noise
generator.



\end{thebibliography}
\end{document}